# PECULIARITIES OF ELECTRONIC TRANSPORT AND MAGNETIC STATE IN HALF-METALLIC FERROMAGNETIC AND SPIN GAPLESS SEMICONDUCTING HEUSLER ALLOYS

V. V. Marchenkov[a,b,*], V. Yu. Irkhin[a,b], Yu. A. Perevozchikova[a]

[a]M.N. Miheev Institute of Metal Physics of UB RAS, 620108 Ekaterinburg, S. Kovalevskaya, 18, Russia
[b]Ural Federal University, 620002 Ekaterinburg, Mira, 19, Russia
*e-mail: march@imp.uran.ru

A brief survey of experimental and theoretical studies of half-metallic ferromagnets (HMFs) and spin gapless semiconductors is given, the possible candidates being the $X_2YZ$ ($X$ = Mn, Fe, Co; $Y$ = Ti, V, Cr, Mn, Fe, Co, Ni; $Z$ = Al, Si, Ga, Ge, In, Sn, Sb) Heusler alloys. The data on the electrical resistivity, normal and anomalous Hall Effect, and magnetic properties are presented. It is shown that the $Co_2FeZ$ alloys demonstrate properties of conventional ferromagnets, the HMF properties being also manifested at the variation of the $Z$-component. The $Fe_2YAl$ and $Mn_2YAl$ alloys show at the variation of the $Y$-component both metallic and semiconducting electronic characteristics, the magnetic properties, changing from the ferromagnetic to compensated ferrimagnetic state. The HMF and spin gapless semiconductor states are supposed to exist in these Heusler alloys systems.

## INTRODUCTION

Search and study half-metallic ferromagnets (HMFs) [1, 2] and spin gapless semiconductors (SGSs) [2, 3] are of great physical and practical interest, since these materials can be used in spintronics [4].

The main feature of the electronic structure of HMFs is the presence of an energy gap at the Fermi level $E_F$ in one spin sub-band and a metallic character of the density of states $N(E)$ in the other [1, 2]. This can lead to 100% spin polarization of the charge carriers.

SGSs were predicted in 2008 [3] as a new class of materials with zero gap for one of spin directions. These compounds should have a number of unique properties associated with their unusual band structure. Such materials make it possible to combine the properties of HMF with semiconducting characteristics and a permit a fine regulation of the energy gap, and hence control of electrical properties.

According to the mean-field picture, HMF can be regarded as a system of two parallel-connected conductors. One of them is a subsystem of current carriers with spin up, and the other is a subsystem with spin down. The first one, the spin up subsystem, has a typical "metallic" conductivity, so that its resistance increases according to a power law with temperature. The spin down subsystem has a "semiconductor" conductivity, i.e. its resistance (conductivity) should decrease (increase) with increasing temperature.

The mean-field picture is considerably modified when including non-quasiparticle (NQP) states owing to electron-magnon scattering [1, 5]. These states violate 100% spin polarization of current carriers: the yield the density of incoherent states in the energy gap, proportional to $T^{3/2}$ (Fig.1). Recently, HMF behavior and presence of NQP states was directly observed in $CrO_2$ by bulk-sensitive spin-resolved photoemission spectroscopy [6].

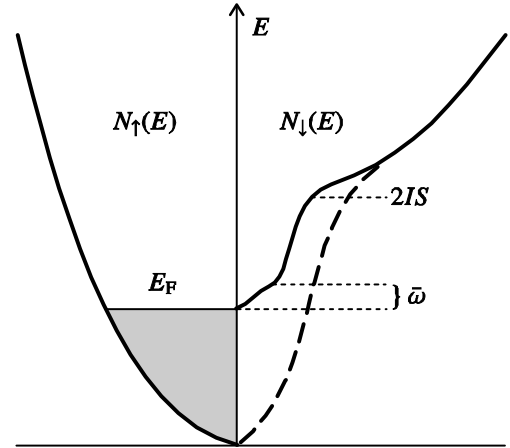

Fig. 1. The density of states in HMF, $2IS$ being the spin splitting, $\varpi$ the characteristic magnon frequency. The dashed line shows the finite temperature tail of NQP states which is proportional to $T^{3/2}$.

The NQP states make considerable contributions to magnetic and transport properties [1,5]. Spin transport in HMF at finite temperatures was theoretically investigated with account of minority-spin band NQP states in Ref. [7]. Spin Hall conductivity proportional to $T^{3/2}$ was found.

The magnetic resistivity of usual ferromagnetic metals is determined by one-magnon processes which yield

$$\rho(T) \sim T^2 N_\uparrow(E_F) \cdot N_\downarrow(E_F) \cdot \exp(-T/T^*) \quad (1)$$

where $T^* \sim q_1^2 T_C$ is the characteristic scale for these processes, $q_1 \sim \Delta/v_F$, $\Delta = 2IS$ being the spin splitting (the energy gap). This contribution is absent for HMF since $N_\downarrow(E_F) = 0$.

Two-magnon scattering processes [8] lead to a power-law temperature dependence of the resistivity $\rho(T) \sim T^n$, as well as to a negative linear magnetoresistance. We have $n = 9/2$ at $T<T^{**}$ and $n = 7/2$ at $T>T^{**}$, the corresponding scale $T^{**}$ being estimated as $T^{**} \sim q_2^2 T_C$. In the simple one-band model of HMF where $E_F < \Delta$ one has $q_2 \sim (\Delta/W)^{1/2}$ with $W$ the bandwidth. Generally speaking, $q_2$ may be sufficiently small provided that the energy gap is much smaller than $W$, which is typical for real HMF systems.

Since many Heusler alloys (HAs) with the general formula $X_2YZ$ ($X$, $Y$ are 3d-elements and $Z$ is s-, p-elements of the Periodic Table) belong to the HMF and SGS materials [4], the study of the electronic structure and magnetic state of such HAs is highly promising. The position and the width of the energy gap can vary quite strongly in different HAs. These parameters can be changed by varying the 3d-, s- and p-elements in $X_2YZ$ Heusler alloys, altering thereby electronic properties.

It is known that HAs Co$_2$FeSi [9], Co$_2$MnSi [10] and Fe$_2$MeAl (Me = V, Ti, Cr, Mn, Fe, Co, Ni) [11, 12] demonstrate HMF properties including a high degree of spin polarization of the current carriers [9, 10]. According to [13], the compound Mn$_2$CoAl exhibits unusual electronic and magnetic characteristics inherent in SGS. Apparently, the peculiarities of HMF and SGS materials will be observed in other alloys of these HA systems, e.g., Co$_2$FeZ, Co$_2$MnZ, Fe$_2$YAl and Mn$_2$YAl, which should manifest themselves in the electrical and magnetic properties. Thus, the purpose of the present work is a brief survey and analysis of the papers on the electron transport and magnetic characteristics of $X_2YZ$ alloys ($X$ = Mn, Fe, Co; $Y$ = Ti, V, Cr, Mn, Fe, Co, Ni; $Z$ = Al, Si, Ga, Ge, In, Sn, Sb) and the establishment of the basic laws of their behavior when varying the $Y$- and $Z$-components.

## EXPERIMENTAL

Synthesis of polycrystalline alloys and their heat treatment are described in detail in [14-16]. Elemental analysis was carried out by using a scanning electron microscope equipped with an EDAX X-ray microanalysis attachment. The deviation from a stoichiometric composition was revealed to be insignificant in all samples. The structural analysis was performed at the Collaborative Access Center, M.N. Miheev Institute of Metal Physics. Methods of measuring electrical resistivity, magnetization, and Hall Effect are presented in [15, 17-19].

## RESULTS ON MAGNETIC AND TRANSPORT PROPERTIES

Figure 2 shows the temperature dependences of the electrical resistivity $\rho(T)$ of the studied alloys. It can be seen that in the Co$_2$FeZ system (Fig. 2a) all compounds have relatively small residual resistivity $\rho_0$, and the dependence $\rho(T)$ is typical for metals, i.e. it monotonically increases with temperature and in the region of the Curie temperature $T_C$ there are features in the form of kinks. The systems Fe$_2$YAl (Fig. 2b) and Mn$_2$YAl (Fig. 2c) have a number of peculiar features. First, the residual resistivity of all alloys, with the exception of the Fe$_2$TiAl compound, is rather large and reaches 2000 μΩ·cm in Fe$_2$VAl. Secondly, for many of them there are wide regions with a negative temperature coefficient of resistance (TCR).

**Table 1.** The saturation magnetization $M_S$, the normal $R_0$ and anomalous $R_S$ Hall coefficients, the residual resistivity $\rho_0$, the current carrier concentration $n$, mobility $\mu$ and the Curie temperatures $T_C$ of Co$_2$FeZ ($Z$ = Al, Si, Ga, Ge, In, Sn, Sb)

| Alloy | $M_S$, emu/g | $R_0 \cdot 10^6$, cm$^3$/C | $R_S \cdot 10^3$, cm$^3$/C | $\rho_0$, μΩ·cm | $n \cdot 10^{-21}$, cm$^{-3}$ | $\mu$, cm$^2$/(s·V) | $T_C$ [14], K |
|---|---|---|---|---|---|---|---|
| Co$_2$FeAl | 150 | -137 | 2.57 | 49.8 | 45.3 | 2.8 | >1100 |
| Co$_2$FeSi | 160 | 70 | 0.22 | 10 | 88.9 | 7 | 1030 |
| Co$_2$FeGa | 146 | -89 | 0.27 | 9.3 | 70.3 | 9.6 | 1056 |
| Co$_2$FeGe | 164 | 36 | 0.21 | 14.4 | 172.6 | 2.5 | 1060 |
| Co$_2$FeIn | 224 | -69 | 0.09 | 1.6 | 90.6 | 43.8 | 890 |
| Co$_2$FeSn | 97 | -87 | 0.34 | 19.7 | 71.5 | 4.4 | 968 |
| Co$_2$FeSb | 99 | -67 | 0.32 | 9.1 | 93.3 | 7.3 | >1100 |

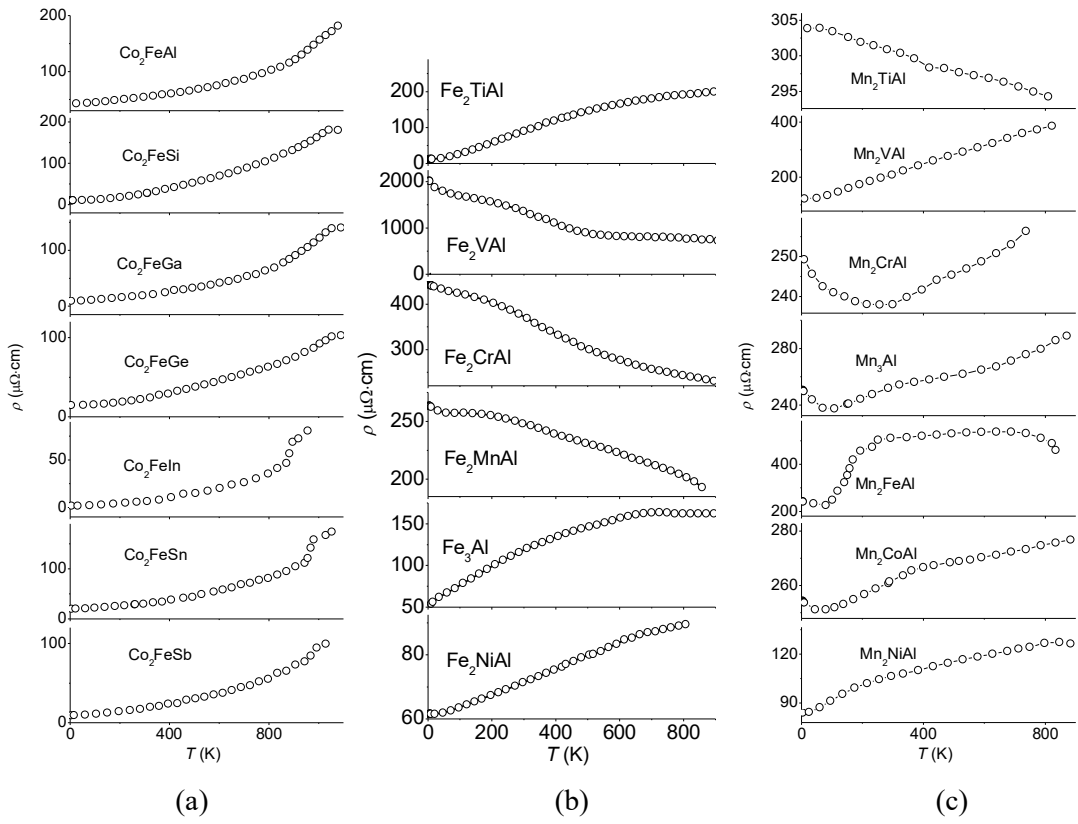

Fig. 2. Temperature dependence of the resistivity of $Co_2FeZ$ (a), $Fe_2YAl$ (b) and $Mn_2YAl$ (c) Heusler alloys, where $Y$ = Ti, V, Cr, Mn, Fe, Co, Ni; $Z$ = Al, Si, Ga, Ge, In, Sn, Sb.

As discussed in the Introduction, in HMF one-magnon scattering processes are suppressed, and two-magnon scattering processes are important, leading to a specific contribution to the temperature dependence of the electrical resistivity of the form $T^n$, where $7/2 < n < 9/2$. The contribution to the conductivity from current carriers with spin up is predominant, and one has in the experiment [9, 14, 20] a good metallic dependence $\rho(T)$ and small $\rho_0$ values, similar to the dependences for $Co_2FeZ$ alloys in Fig. 2a.

A small (or nearly zero) gap near the Fermi energy $E_F$ for both spin projections can lead to regions with negative TCR and large $\rho_0$ values, as is observed in some alloys of $Fe_2YAl$ and $Mn_2YAl$ systems (Fig. 2b and 2c). It can be assumed that the presence of a gap for one or two projections of the spin can also manifest itself in other transport, as well as magnetic properties. This should be especially noticeable at temperatures much lower than $T_C$. Therefore, the magnetization and the Hall Effect at the liquid helium temperature $T = 4.2$ K were studied.

Fig. 3 shows the field dependence of the magnetization of the alloy systems $Co_2FeZ$, $Fe_2YAl$ and $Mn_2YAl$ at $T = 4.2$ K. The magnetization of $Co_2FeZ$, $Fe_2YAl$, $Mn_2VAl$ and $Mn_2CoAl$ is seen to has saturation at high fields ($H > 10$ kOe). The saturation magnetization $M_S$ of these alloys was determined, as can be seen in Tables 1-3. The dependences of the Hall resistivity $\rho_H(H)$ (Fig. 4) behave like the field dependences of the magnetization (Fig. 3). For all alloys studied, the Hall coefficients were determined. In the case of ferromagnetic alloys, the normal $R_0$ and anomalous $R_S$ Hall coefficients were picked up taking into account [15] and using the expression

**Table 2**. The saturation magnetization $M_S$, the normal $R_0$ and anomalous $R_S$ Hall coefficients, the residual electroresistivity $\rho_0$, the current carrier concentration $n$, mobility $\mu$ and the Curie temperatures $T_C$ of $Fe_2YAl$ ($Y$ = Ti, V, Cr, Mn, Fe, Ni)

| Alloy | $M_S$, emu/g | $R_0 \cdot 10^4$, cm$^3$/C | $R_S$, cm$^3$/C | $\rho_0$, µΩcm | $n \cdot 10^{-21}$, cm$^{-3}$ | $\mu$, cm$^2$/(s·V) | $T_C$ [11], K |
|---|---|---|---|---|---|---|---|
| $Fe_2TiAl$ | 0.07 | 2.3 | - | 11.5 | 27.3 | 19.9 | 123 |
| $Fe_2VAl$ |  | 458 | - | 2020 | 0.14 | 22.7 | 7 |
| $Fe_2CrAl$ | 0.05 | -17.2 | 0.54 | 443 | 3.63 | 3.88 | 246 |
| $Fe_2MnAl$ | 0.24 | -15.3 | 0.11 | 263 | 4.1 | 5.82 | 150 |
| $Fe_2FeAl$ | 0.11 | -2.1 | 0.05 | 53 | 29.6 | 3.98 | 775 |
| $Fe_2NiAl$ | 0.21 | -5.4 | 0.02 | 61.5 | 11.5 | 8.8 |  |

$$\rho_H / H = R_0 + 4\pi R_S M / H \qquad (2)$$

where $M$ is the magnetization. The first term $R_0$ describes the normal Hall Effect associated with the action of the Lorentz force on the movement of conduction electrons in a magnetic field $H$. The second term is connected with spin-orbit interaction.

As a result of the magnetization and Hall Effect studies, the values of the saturation magnetization $M_S$, the normal $R_0$ and anomalous $R_S$ Hall coefficients were determined, and the current carrier concentration and mobility of all the studied alloys were estimated (Tables 1-3). It can be seen (see Table 1) that the $Co_2FeZ$ system alloys have low residual resistivity and relatively high saturation magnetization values. The values of the anomalous Hall coefficient $R_S$ are several orders of magnitude greater than the values of the normal Hall coefficient $R_0$, the carrier concentration values being typical for metals.

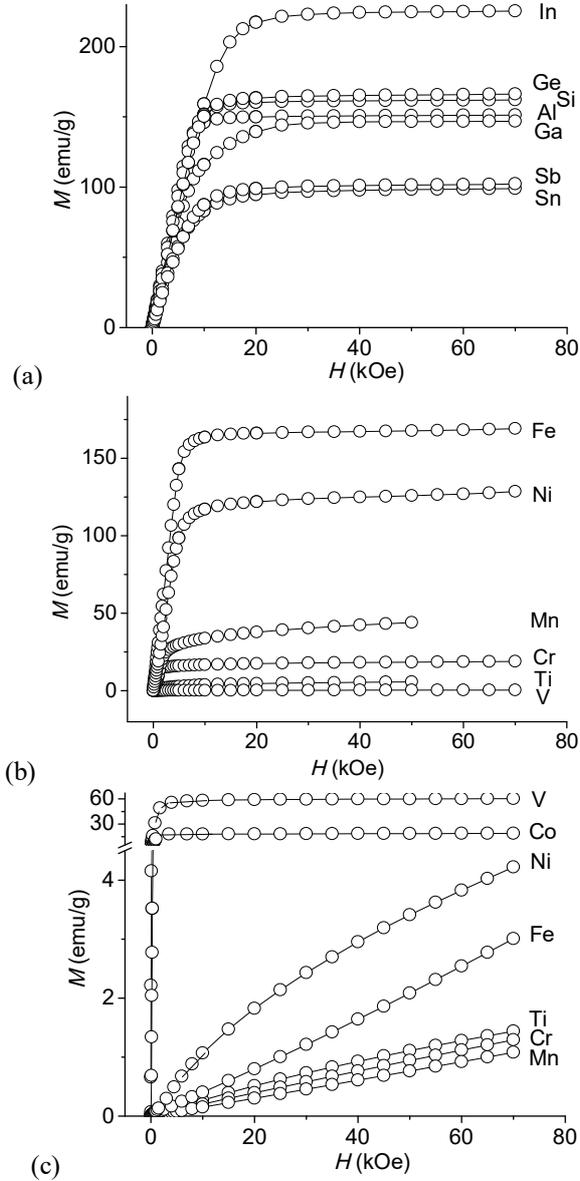

Fig. 3. Field dependence of the magnetization of $Co_2FeZ$ (a), $Fe_2YAl$ (b) and $Mn_2YAl$ (c) Heusler alloys at $T = 4.2$ K. Here $Y$ = Ti, V, Cr, Mn, Fe, Co, Ni; $Z$ = Al, Si, Ga, Ge, In, Sn, Sb.

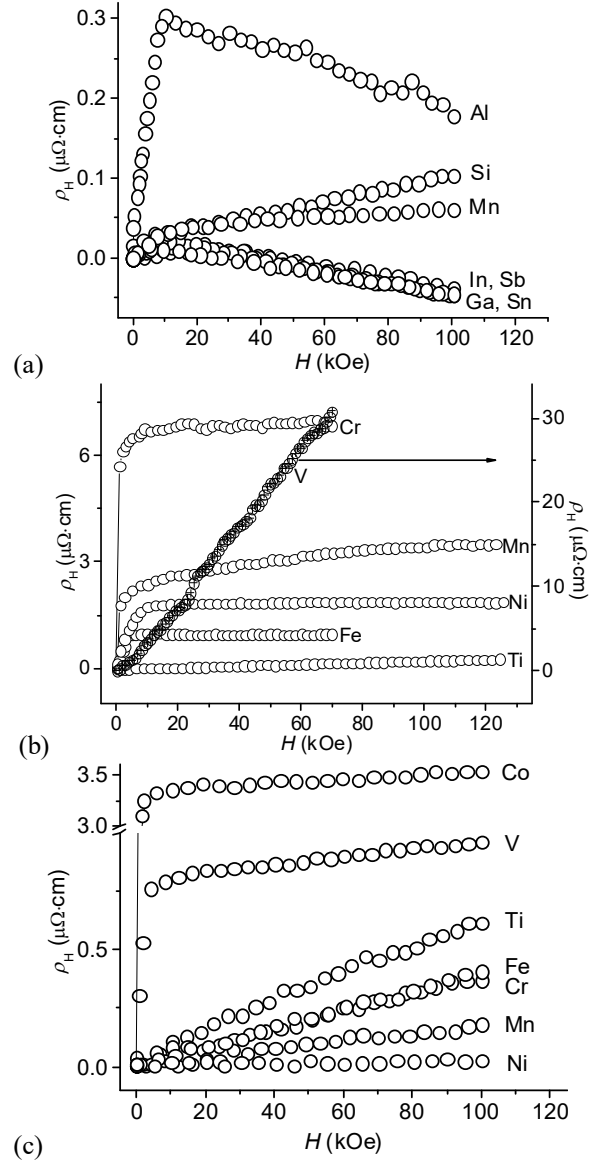

Fig. 4. Field dependence of the Hall resistivity $\rho_H$ of $Co_2FeZ$ (a), $Fe_2YAl$ (b) and $Mn_2YAl$ (c) Heusler alloys at $T = 4.2$ K. Here $Y$ = Ti, V, Cr, Mn, Fe, Co, Ni; $Z$ = Al, Si, Ga, Ge, In, Sn, Sb.

In the case of the $Fe_2YAl$ system (see Table 2), both relatively small ($Fe_2TiAl$, $Fe_3Al$, $Fe_2NiAl$) and enormous ($Fe_2VAl$) residual resistivity values, along with the "metallic" behavior $\rho(T)$, the presence of wide temperature regions with a negative temperature coefficient of resistance (TCR), the magnetic state of the alloys varies from weakly magnetic ($Fe_2TiAl$, $Fe_2VAl$) to "good" ferromagnets ($Fe_3Al$, $Fe_2NiAl$), and the values of carrier concentrations vary from "semiconductor" values ($Fe_2VAl$) to those typical for "bad" metals ($Fe_2TiAl$, $Fe_2FeAl$, $Fe_2NiAl$).

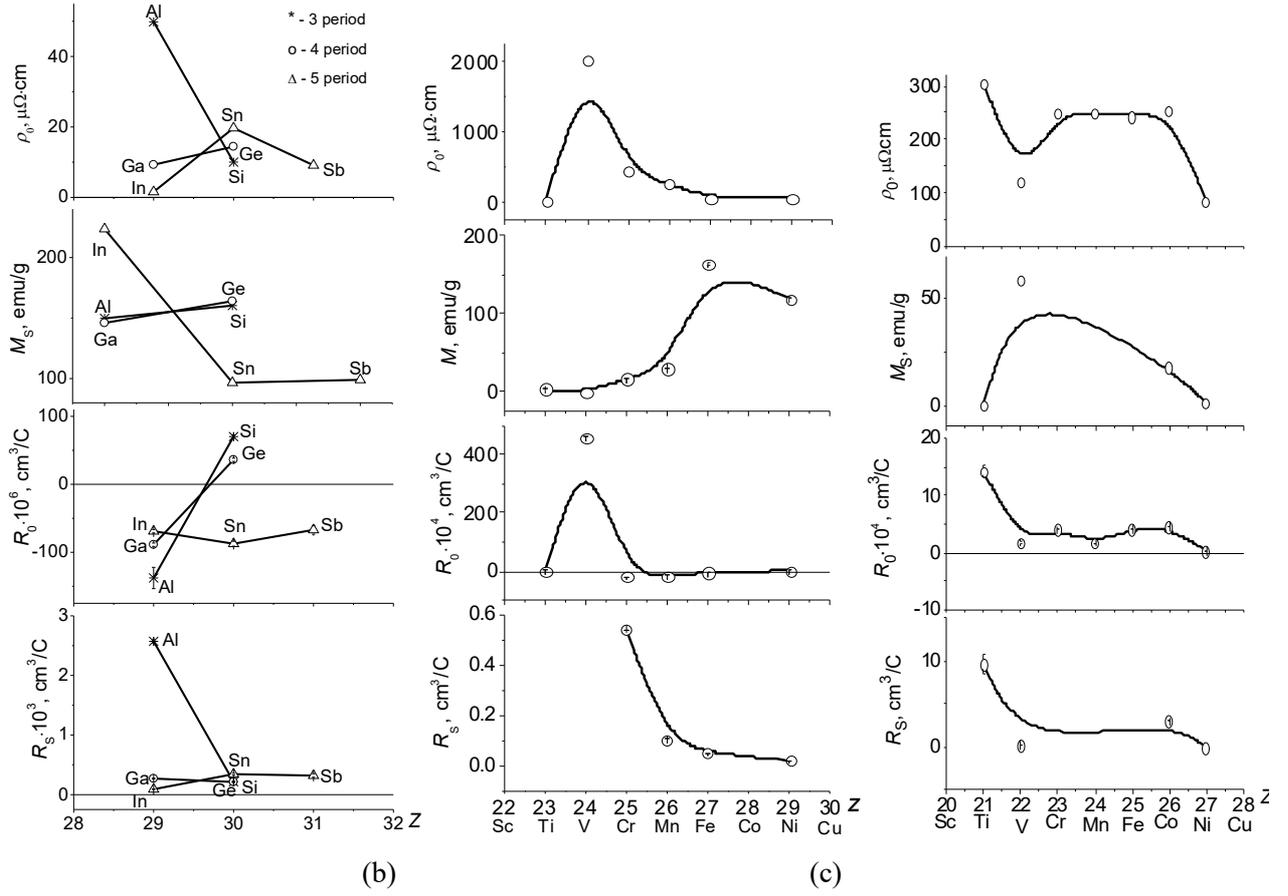

(b)    (c)

Fig. 5. The residual resistivity $\rho_0$, the saturation magnetization $M_S$, the coefficients of normal $R_0$ and anomalous $R_S$ Hall Effects for $Co_2FeZ$ (a), $Fe_2YAl$ (b) and $Mn_2YAl$ (c).

Finally, for $Mn_2YAl$ system (see Table 3), a relatively large residual resistivity $\rho_0$, with the exception of the $Mn_2VAl$ and $Mn_2NiAl$ alloys, for which the "metallic" behavior of $\rho(T)$ is observed. For other compounds of this system there are regions with negative TCR. Contrary to the electronic structure calculations [21], magnetic measurements for the $Mn_2CrAl$, $Mn_3Al$ and $Mn_2FeAl$ systems give zero total magnetization and may indicate compensated ferrimagnetism [22, 23].

In order to treat the changes in the above mentioned electronic and magnetic characteristics and to establish a possible correlation between them at varying $Y$ and $Z$ components, the dependences of $\rho_0$, $M_S$, $R_0$ and $R_S$ on the number of valence electrons $z$ were plotted (Fig. 5). A distinct correlation is seen between these quantities for all the alloys. For $Fe_2YAl$ and $Mn_2YAl$ alloys (Fig. 5b and 5c), the maximum (minimum) of the residual resistivity corresponds to the minimum (maximum) of magnetization and the maximum (minimum) of the normal Hall coefficient. As already noted in [24], the number of valence electrons $z$ is a rather crude characteristic which does not represent properly peculiarities of a particular compound in terms of its electronic band structure features. The density of electron states $N(E)$ at the Fermi level $E_F$ could be such a parameter. Apparently, it would be more correct to present the dependences of the electronic and magnetic characteristics shown in Fig. 5 as a function of $N(E_F)$, which could allow a future quantitative description of the correlation between the electronic and magnetic characteristics. For Co$_2$Fe$Z$ ($Z$ = Al, Si, Ga, Ge) alloys, the correlation is seen to occur too. Namely, this is the change of main current carrier type from holes to electrons (i.e., sign change of normal Hall coefficient $R_0$), the increase of saturation magnetization $M_S$ and decrease of the anomalous Hall coefficient $R_S$ at transition from third to fourth period of the Periodic Table.

**Table 3.** The saturation magnetization $M_S$, the normal $R_0$ and anomalous $R_S$ Hall coefficients, the residual electroresistivity $\rho_0$, the current carrier concentration $n$, mobility $\mu$ and the Curie temperatures $T_C$ of $Mn_2YAl$ ($Y$ = Ti, V, Cr, Mn, Fe, Co, Ni)

| Alloy | $M_S$, emu/g | $R_0·10^4$, cm$^3$/C | $R_S$, cm$^3$/C | $\rho_0$, $\mu\Omega$cm | $n·10^{-21}$, cm$^{-3}$ | $\mu$, cm$^2$/(s·V) | $T_C$ [27], K |
|---|---|---|---|---|---|---|---|
| $Mn_2TiAl$ | 0.2 | 14.2 | 9.67 | 305.5 | 4.4 | 4.6 | 428 |
| $Mn_2VAl$ | 58.9 | 1.76 | 0.24 | 122.3 | 35.4 | 1.4 | 588 |
| $Mn_2CrAl$ | - | 4 | - | 250.4 | 15.6 | 1.6 | 549 |
| $Mn_2MnAl$ | - | 1.71 | - | 250.6 | 36.5 | 0.7 | 196 |
| $Mn_2FeAl$ | - | 4.06 | - | 242.1 | 15.4 | 1.7 | 200 |
| $Mn_2CoAl$ | 17.8 | 4.46 | 3.02 | 254.3 | 14 | 1.8 | 740 |
| $Mn_2NiAl$ | 1.2 | 0.18 | 0 | 84.1 | 34.7 | 0.2 | 452 |

CONCLUSIONS

As a result of studying the electron transport and magnetic properties in $X_2YZ$ alloys ($X$ = Mn, Fe, Co; $Y$ = Ti, V, Cr, Mn, Fe, Co, Ni; $Z$ = Al, Si, Ga, Ge, In, Sn, Sb) it was found that at varying the $Y$ and $Z$ components the systems can vary from usual metals to half-metallic systems and, possibly, spin gapless semiconductors. Thus we can conclude.

1. In the Co$_2$Fe$Z$ system alloys, at the variation of the $Z$-component one observes:

- small values of residual resistivity $\rho_0$ and metallic type of temperature dependences $\rho(T)$;

- all alloys are ferromagnets with relatively high values of magnetization and Curie temperatures $T_C$;

- the values of the anomalous Hall coefficient $R_S$, are, as a rule, several orders of magnitude greater than the values of the normal Hall coefficient $R_0$;

- values of carrier concentrations are typical for metals;

- the contribution to the resistivity is proportional to $T^n$, $7/2 < n < 9/2$, which may indicate two-magnon scattering processes; So, in addition to the properties of conventional ferromagnets, the HMF properties are also manifested.

2. In the Fe$_2Y$Al alloys, at the variation of the $Y$-component one has:

- both relatively small (Fe$_2$TiAl, Fe$_3$Al, Fe$_2$NiAl) and enormous (Fe$_2$VAl) residual resistivity values, along with the "metallic" behavior $\rho(T)$, the presence of wide temperature regions with a negative temperature coefficient of resistance (TCR).

- The magnetic state of the alloys varies from weakly magnetic (Fe$_2$TiAl, Fe$_2$VAl) to "good" ferromagnets (Fe$_3$Al, Fe$_2$NiAl);

- values of carrier concentrations vary from "semiconductor" values (Fe$_2$VAl) to values, typical for "bad" metals (Fe$_2$TiAl, Fe$_2$FeAl, Fe$_2$NiAl);

- in addition to the properties of a conventional ferromagnetic metal, the HMF properties also appear. The Fe$_2$VAl alloy demonstrates nearly semiconducting properties. According to [25, 26], this system exhibits the transition from a semiconducting to metallic state with the onset of ferromagnetic order. Thus, in the Fe$_2Y$Al alloys, peculiarities of the electron energy spectrum near the Fermi level $E_F$ play an important role in electronic transport.

3. In the Mn$_2Y$Al alloys, at varying the $Y$-component one observes:

- a relatively large residual resistivity $\rho_0$, with the exception of the Mn$_2$VAl and Mn$_2$NiAl alloys, for which the "metallic" behavior of $\rho(T)$ is observed. For other compounds of this system there are regions with negative TCR;

- contrary to the electronic structure calculations, magnetic measurements for the Mn$_2$CrAl, Mn$_3$Al and Mn$_2$FeAl systems give zero total magnetization and may indicate compensated ferrimagnetism.

The anomalies found can indicate peculiarities of the electron energy spectrum: the appearance of states of a half-metallic ferromagnet or spin gapless semiconductor. The results obtained may be of interest for the development of new materials applicable to spintronics.

This work was partly supported by the Ministry State Assignment of Russia (themes "Spin" No. AAAA-A18-118020290104-2 and "Quant" No. AAAA-A18-118020190095-4), RFBR grant (No. 18-02-00739, 18-32-00686) and the Government of the Russian Federation (state contract No. 02.A03.21.0006).

REFERENCES

1. V. Y. Irkhin, and M. I. Katsnelson, "Half-metallic ferromagnets," Physics Uspekhi. **37,** 659 (1994).
2. V. V. Marchenkov, N. I. Kourov, and V. Yu. Irkhin, "Half-metallic ferromagnets and spin gapless semiconductors," Phys. Met. Metallogr. **119**, 1321 (2018).
3. X. L. Wang, "Proposal for a new class of materials: Spin gapless semiconductors," Phys. Rev. Lett. **100**, 156404 (2008).
4. T. Graf, C. Felser, and S. S. P. Parkin, "Simple rules for the understanding of Heusler compounds," Prog. Solid State Chem. **39**, 1 (2011).
5. V. Yu. Irkhin, M. I. Katsnelson, and A. I. Lichtenstein, "Non-quasiparticle effects in

half-metallic ferromagnets," J. Phys.: Condens. Matter **19**, 315201 (2007).
6. H. Fujiwara, M. Sunagawa, K. Terashima, T. Kittaka, T. Wakita, Y. Muraoka, and T. Yokoya, "Observation of Intrinsic Half-metallic Behavior of $CrO_2$ (100) Epitaxial Films by Bulk-sensitive Spin-resolved PES," J. Electron Spectrosc. Relat. Phenom. **220**, 46 (2017).
7. Y. Ohnuma., M. Matsuo, and S. Maekawa, "Spin transport in half-metallic ferromagnets," Phys. Rev. B. **94**, 184405 (2016).
8. V. Yu. Irkhin, and M. I. Katsnelson, "Temperature dependences of resistivity and magnetoresistivity for half-metallic ferromagnets," Eur. Phys. J. B **30**, 481 (2002).
9. D. Bombor, C. G. F. Blum, O. Volkonskiy, S. Rodan, S. Wurmehl, C. Hess, and B. Buchner, "Half-Metallic Ferromagnetism with Unexpectedly Small Spin Splitting in the Heusler Compound $Co_2FeSi$," Phys. Rev. Lett. **110**, 066601 (2013).
10. M. Jourdan, J. Minar, J. Braun, A. Kronenberg, S. Chadov, B. Balke, A. Gloskovskii, M. Kolbe, H. J. Elmers, G. Schonhense, H. Ebert, C. Felser, and M. Klaui, "Direct observation of half-metallicity in the Heusler compound $Co_2MnSi$," Nat. Commun. **5**, 3974 (2014).
11. N. I. Kourov, V. V. Marchenkov, K. A. Belozerova, and H. W. Weber, "Specific features of the electrical resistivity of half-metallic ferromagnets $Fe_2MeAl$ (Me = Ti, V, Cr, Mn, Fe, Ni)," J. Exp. Theor. Phys. **118**, 426 (2014).
12. N. I. Kourov, V. V. Marchenkov, A. V. Korolev, L. A. Stashkova, S. M. Emel'yanova, and H. W. Weber, "Specific features of the properties of half-metallic ferromagnetic Heusler alloys $Fe_2MnAl$, $Fe_2MnSi$, and $Co_2MnAl$," Phys. Solid State **57**, 700 (2015).
13. S. Ouardi, G. H. Fecher, C. Felser, and J. Kubler, "Realization of Spin Gapless Semiconductors: The Heusler Compound $Mn_2CoAl$," Phys. Rev. Lett. **110**, 100401 (2013).
14. V. V. Marchenkov, Yu. A. Perevozchikova, N. I. Kourov, V. Yu. Irkhin, M. Eisterer, and T. Gao, "Peculiarities of the electronic transport in half-metallic Co-based Heusler alloys," J. Magn. Magn. Mater. **459,** 211 (2018).
15. N. I. Kourov, V. V. Marchenkov, K. A. Belozerova, and H. W. Weber, "Galvanomagnetic properties of $Fe_2YZ$ (Y = Ti, V, Cr, Mn, Fe, Ni; Z = Al, Si) Heusler alloys," J. Exp. Theor. Phys. **121**, 844 (2015).
16. N. I. Kourov, V. V. Marchenkov, A. V. Korolev, K. A. Belozerova, and H. W. Weber, "Peculiarities of the electronic transport in $Co_2CrAl$ and $Co_2CrGa$ half-metallic ferromagnets," Current Applied Physics **15**, 839 (2015).
17. N. V. Volkenshtein, V. V. Marchenkov, V. E. Startsev, A. N. Cherepanov, and M. Glin'ski, "Hall Effect accompanying a static skin effect," JETP Lett. **41,** 458 (1985).
18. N. V. Volkenshtein, M. Glin'ski, V. V. Marchenkov, V. E. Startsev, and A. N. Cherepanov, "Characteristics of galvanomagnetic properties of compensated metals under static skin effect conditions in strong magnetic fields (tungsten)," J. Exp. Theor. Phys. **68**, 1216 (1989).
19. A. N. Cherepanov, V. V. Marchenkov, V. E. Startsev, N. V. Volkenshteyn, and M. Glinskii, "High-field galvanomagnetic properties of compensated metals under electron-surface and intersheet electron-phonon scattering (tungsten)," J. Low Temp. Phys. **80**, 135 (1990).
20. K. Srinivas, M. Manivel Raja, and S. V. Kamat, "Effect of partial substitution of silicon by other sp-valent elements on structure, magnetic properties and electrical resistivity of $Co_2FeSi$ Heusler alloys," J. Alloys Comp. **619**, 177 (2015).
21. H.Z. Luo, Z.Y. Zhu, L. Ma, S.F. Xu, X.X. Zhu, C.B. Jiang, H.B. Xu, and G.H. Wu, "Effect of site preference of 3d atoms on the electronic structure and half-metallicity of Heusler alloy $Mn_2YAl$," J. Phys. D: Appl. Phys. **41**, 055010 (2008).
22. M. E. Jamer, Y. J. Wang, G. M. Stephen, I. J. McDonald, A. J. Grutter, G. E. Sterbinsky, D. A. Arena, J. A. Borchers, B. J. Kirby, L. H. Lewis, B. Barbiellini, A. Bansil, and D. Heiman, "Compensated ferrimagnetism in the zero-moment Heusler alloy $Mn_3Al$," Phys. Rev. Appl. **7**, 064036 (2017).
23. V. V. Marchenkov, V. Yu. Irkhin, Yu. A. Perevozchikova, P. B. Terent'ev, A. A. Semiannikova, E. B. Marchenkova, and M. Eisterer, "Kinetic properties and half-metallic magnetism in $Mn_2Y$Al Heusler alloys," J. Exp. Theor. Phys. **128**, 919 (2019).
24. Yu. A. Perevozchikova, A. A. Semiannikova, A. N. Domozhirova, E. I. Patrakov, M. Eisterer, P. S. Korenistov, and V. V. Marchenkov, "Experimental observation of anomalies in the electrical, magnetic, and galvanomagnetic properties of cobalt-based Heusler alloys with varying transition elements," Low Temp. Phys. **45**, 015907 (2019).
25. Y. Nishino, M. Kato, S. Asano, K. Soda, M. Hayasaki, and U. Mizutani, "Semiconductorlike behavior of electrical resistivity in Heusler-type $Fe_2VAl$ compound," Phys. Rev. Lett. **79**, 1909 (1997).

26. V. I. Okulov, A. T. Lonchakov, and V. V. Marchenkov, "Semiconductor-like behavior of electric transport in Fe-V-Al-based metallic alloys and their uncommon magnetic properties," Phys. Met. Metallogr. **119**, 1325 (2018).

27. L. Wollmann, S. Chadov, J. Kübler, and C. Felser, "Magnetism in cubic manganese-rich Heusler compounds," Phys. Rev. B **90**, 214420 (2014).